\documentclass[twocolumn,prb,showpacs,amsmath,amssymb]{revtex4}

\usepackage{amsmath,amsfonts,amsthm,graphicx,color}
\usepackage{epsfig}
\usepackage{float}
\usepackage{hyperref}

\begin{document}

\title{Large Berry phases in layered graphene}
\author{R. Englman}
\email{englman@vms.huji.ac.il}
\affiliation{Soreq NRC, Yavne 81800, Israel}
\author{T. V\'ertesi}
\email{tvertesi@dtp.atomki.hu}
\affiliation{Institute of Nuclear Research of the Hungarian Academy of Sciences\\
H-4001 Debrecen, P.O. Box 51, Hungary}

\def\CC{\mathbb{C}}
\def\RR{\mathbb{R}}
\def\one{\leavevmode\hbox{\small1\normalsize\kern-.33em1}}
\newcommand*{\tr}{\mathsf{Tr}}
\newtheorem{theorem}{Theorem}[section]
\newtheorem{lemma}[theorem]{Lemma}

\newcommand{\beq} {\begin{equation}}
\newcommand{\enq} {\end{equation}}
\newcommand{\ber} {\begin {eqnarray}}
\newcommand{\enr} {\end {eqnarray}}
\newcommand{\eq} {equation}
\newcommand{\eqs} {equations }
\newcommand{\mn}  {{\mu \nu}}
\newcommand{\sn}  {{\sigma \nu}}
\newcommand{\rhm}  {{\rho \mu}}
\newcommand {\SE} {Schr\"{o}dinger equation}
\newcommand{\sr}  {{\sigma \rho}}
\newcommand{\bh}  {{\bar h}}
\newcommand {\er}[1] {equation (\ref{#1}) }
\newcommand {\Er}[1] {Equation (\ref{#1}) }
\newcommand {\erb}[1] {equation (\ref{#1})}
\newcommand{\kt} {{\tilde k}}

\date{\today}

\begin{abstract}
Brillouin zones of graphene systems possess Dirac points, where
band degeneracies occur. We study the variety of (and large
magnitude) phases that the electronic states can acquire when a
uniform  time-dependent electric field carries the electrons
around one or more Dirac points in  a non-concentric fashion. An
experimentally accessible determination of excess Berry phases is
proposed involving the Zitterbewegung of electronic current near
an orthogonality point in adiabatic motion.
\end{abstract}

\maketitle


\section{Introduction}
The recent research interest in graphene owes much to the
remarkable electronic properties near the charge-neutrality, Dirac
points (DP's, conical intersections or linear electronic
degeneracies)
  at the ${\bf K}$ and ${\bf K'}$-points in the Brillouin zone~\cite{NovoselovGMJZDGF,NovoselovGMJKGDF,ZhangTSK,Katsnelson,GeimN,NGPNG}. Also several studies devoted to graphene layers of various thicknesses have appeared recently
  ~\cite{NovoselovMMFKZJSG,KoshinoA,PartoensP,McCannF,ManesGV,CsertiCD,MikitikS}. Based on the earlier elucidation of the
   Brillouin-zone structure
    in Refs.~\onlinecite{McClure, SlonczewskiW} and the more recent works in Refs.~\onlinecite {PeresGN}-\onlinecite{MalyshevaO}, in bilayered graphene one centrally
  and three trigonally arranged DP's around the {\bf K}-type points can be located. The electronic motion is described by a Hamiltonian that includes interactions between different
sites in either plane and between the two planes and expressed the coupling between four electronic
bands. An approximate and simplified model using two $2\times2$ Hamiltonians (for the ${\bf K}$ and
${\bf K'}$-points, respectively) was given in Ref.~\onlinecite{McCannF}. The repercussions of
different Hamiltonians on the zero energy minimal conductivity
 were discussed in Ref.~\onlinecite{CsertiCD}, where the effect of the trigonal distortion was found to be significant, whereas the
 simplified Hamiltonian of Ref.~\onlinecite{KoshinoA} led to results similar to those of the four band model.

 Another development
 regarded the Berry phase \cite{Berry, BliokhB} acquired upon circling around the DP's in the ${\bf k}$-plane \cite{NovoselovMMFKZJSG, McCannF, ZhangTSK, MikitikS}. It was assumed in these works that a circling around each point leads to
an added value of $\pm \pi$
 to the geometric phase, with the sign depending on derivatives in the neighborhood of the
 intersection points.
 (The sign of the Berry phase encountered in the molecular physics, electron-vibration context was determined
in Refs.~\onlinecite{EnglmanYDebr, VertesiB}.) It is the purpose  of the
present work to show that when a Berry phase is actually {\it
created} in a physical, adiabatically cyclic process, then the
magnitude of the phase will differ from the assumed value of $\pm
\pi$ and is more correctly $(2n+1) \pi$, where $n$ is a signed
integer or zero. The above result holds for a single DP, when the circling is
performed in a non-concentric manner (and, likewise, when the
circling is around an odd number of intersections). Analogously,
when the circle is around two or an even number of intersection
points the assumed result of $\pm 2\pi$ or zero is
 superseded by $2n\pi$ (with $n$ taking values, as before).  These predictions  are  based
on a proper treatment of the {\it time-dependent} adiabatic
process, which requires near an orthogonality point a formalism
 that involves a correction term beyond the extreme adiabatic limit \cite{EnglmanV1,VertesiE,EnglmanV2}.
(At an "orthogonality point"  the momentary wave function has zero overlap with the starting wave
function. The integer $n$ is related to the varying speed of the electron along its cyclic path. This
is quantified by the parameter $n'$, named "the inverse relative slowness" and  defined below in
equation~(\ref{ntag}). In the present context and formalism the extreme adiabatic limit is defined by
the vanishing of the ratio $\epsilon$ in equation (14) for all times, while n' in equation (18) is
finite and possibly numerically large. A conventional treatment would first neglect the second and
third terms in equation (16) and then obtain a $\pm\pi$ Berry phase from the sign change of the first
term. We retain the remaining terms and calculate the acquired phases at the several zeros near the
orthogonality point, neglecting the small terms only after exiting from the orthogonality
neighborhood.)

 Now it is true that
Berry phases are usually
 written $modulo~ 2\pi$, and are gauge invariant only under conditions of complete cycling, but the effect of the extra $|n|\neq 0$ shows up even before completion of the cycling, as an open path phase \cite{Pati1?}, and can
  in principle be measurable. (The extra acquired phase
 is gauge invariant \cite{EnglmanV1,VertesiE,EnglmanV2}. Experimental verification of the extra open path phase
  has been proposed before \cite{Bhandari}.) An essential
  point (not present in  previous electron-nuclear settings) is that in graphene it is technically feasible
   to control the phase
  acquisition by varying the electron concentration and by an
  electric field. The former is achieved by the manipulation of the (time-constant) gate voltage (through the substrate)
   and the latter by a well designed application of a time-varying uniform in-plane electric field
  that guides the electronic motion through the "acceleration theorem" (of which more in the
  sequel).
    Application of a time independent electric field  was treated
in Ref.~\onlinecite{MartinezJT}. In the present setting of the {\it
adiabatic} change in the electronic wave vector,
 the magnitude of the applied electric field would be of the order of $10^4$ V/m or less; this is  rather less than
 the value $10^7$ V/m  proposed in the preceding reference, appropriate to conditions that are not adiabatic.

   Although in this work we do not detail the
   experimental aspects of the phase acquisition, we do propose and investigate theoretically an entirely new method
   for the experimental observation of the
   Berry phase, feasible for high Berry phase values. (In our future mentioning of the phase, the dynamic phase of Ref.~\onlinecite{Berry} is understood to be
   subtracted from the total wave-function phase. We return to the dynamic phase  in section~\ref{dyn}.) This is made possible
   {\it during} the acquisition of the phase in the vicinity of the
   orthogonality point and is measurable through the Zitterbewegung (ZB) of the electronic motion
   (current). ZB in graphene related materials was the subject of
   several works \cite{Katsnelson2006, RusinZ1, CsertiD, RusinZ2}. (Also related is reference~\onlinecite{SchliemannLW}.)
   It is shown in section~\ref{ZB} that in a adiabatically time developing eigenstate, ZB is
   observable near the orthogonality points. The time duration of
   the ZB is associated with the inverse relative slowness
   parameter. In the present context where adiabatic changes are considered, the duration
   for which ZB has to be observed is expected to be rather large, of the order of $10^3$ fs or more.
   (Restrictions on the duration of the ZB by the finite width of the electronic wave-packet were previously considered in Refs.~\onlinecite{RusinZ1,RusinZ2, Lock}.)

  \section{Graphene-based Hamiltonians}
\subsection{Monolayer graphene}\label{mono}
This is a zero-gap semiconductor, whose zero-temperature
electronic properties in the undoped or slightly doped form arise
from the neighborhood of hexagonally arranged ${\bf K}$-points
(the Dirac points, DP) \cite{GeimN}. Here the valence and
conduction bands meet in a conical intersection.  The Hamiltonian,
expressed by means of the planar quasi-particle momenta $(k_x,
k_y)$ measured from a ${\bf K}$-point, have the form of a
two-dimensional Dirac-Hamiltonian \beq H =v_F\left(
\begin{array}{cc}
   k_x &  k_y\\
  k_y &  -k_x\end{array} \right),
  \label{monoham} \enq where a real representation was adopted for
  the electronic band states and $v_F$ is the Fermi velocity
  (about $10^6$ m/s). The off-diagonal, complex  Hamiltonian commonly used in the graphene
  literature is $M^{\dag}HM$,
   having applied the unitary transformation matrix
   \beq M =\frac{1}{\sqrt{2}}\left(
\begin{array}{cc}
   1 &  1\\
  -i &  i\end{array} \right),
  \label{transfMatr}\enq
   to the above Hamiltonian H.

For future use we note that for momenta $|k|$ whose cyclic time
rate of change is represented by $\omega$, the requirement of
adiabaticity is given in a monolayer graphene by \beq v_F|{\bf
k}|\gg \omega.\label{adiabatic}\enq

  \subsection{Graphene bilayer}
  Four electronic bands describe the salient properties.
  A concise form of the $4$-band Hamiltonian for an electron in a ${\bf k}$ (wave vector) state near the ${\bf K}$
  and ${\bf K'}$-points  was written out in Ref.~\onlinecite{CsertiCD}. (Eq.~(1) there.) An approximate $2$-band Hamiltonian proposed by Ref.~\onlinecite{McCannF} can be written in a real electronic state
  representation and following the real-matrix notation used in Refs.~\onlinecite{EnglmanV1}-\onlinecite{EnglmanV2} as
  \beq H_{\bf K}({\bf \kt}) =\left(
\begin{array}{cc}
   -U({\bf \kt}) &  V({\bf \kt})\\
  V({\bf \kt}) &  U({\bf \kt})\end{array} \right)
  \label{HK} \enq
for (e.g.) the ${\bf K}$-point. Here the designation ${\bf \kt}$
is used for the reduced in-plane wave-vector measured from the
${\bf K}$-point. \beq {\bf \kt}\equiv\frac{\bf k}{k_0}=
\frac{(k_x, k_y)}{k_0}\equiv(\kt_x,\kt_y),\label{kt}\enq where
$k_0=\frac{2\gamma_1\gamma_3} {\sqrt{3}a\gamma_0^2}$ having
introduced the intra-layer coupling strength $\gamma_0$ between
the basis atoms A and B within the first layer and between A' and
B' within the second layer, the vertical interlayer A-B' coupling
strength $\gamma_1$, the (weaker) diagonal interlayer A'-B
coupling strength $\gamma_3$ and the honeycomb lattice constant
{\it a}. Numerical values that have been accepted for these
parameters are $\gamma_0$=3.16 eV, $\gamma_1$=0.39 eV,
$\gamma_3$=0.315 eV, $a$= 0.246nm, $k_0$= 0.05775 nm$^{-1}$. The
matrix elements
are \ber U({\bf \kt}) & =  & e_0[\kt_x-(\kt_x^2-\kt_y^2)],\label{U}\\
V({\bf \kt}) & = & e_0[\kt_y+2 \kt_x\kt_y]\label {V},\enr
with an overall coupling energy  $e_0\equiv\gamma_1 (\frac{\gamma_3}{\gamma_0})^2= 3.87$ meV. For a ${\bf K'}$-point
 the signs of the first term in $U({\bf \kt})$ and of the second term in $V({\bf \kt})$ need to be changed.
 \section{Motion of electrons}
 When subject to an electric field $\bf E$, the wave vector of an electron (with charge $-e$) changes in time ($t$)
 according to
 the "acceleration theorem" \cite{LandauL,ZakSSP} \beq\frac{d{\bf k}}{dt}=-\frac{e}{\hbar}{\bf E}.\label{kdot}\enq
A spatially uniform and time  varying electric field is expressed
in terms of a vector potential ${\bf A}(t)$ through \beq {\bf
E}=-\frac{\partial {\bf A}(t)}{\partial t}\label{E}\enq giving
\beq{\bf k}(t)= \frac{e}{\hbar}{\bf A}(t)\label{k(t)}\enq
 apart from an initial value of the wave vector.
Thus, the above Hamiltonians are to be understood as functions of
the vector potential or of the time integral of the externally
applied electric field. This will be the meaning attached
 to the independent variables (${\bf k}$ or ${\bf \kt}$) in the Hamiltonians, though (for
simplicity of notation) we shall continue to write them as functions
of the time-dependent
 planar wave-vectors,  rather than that of ${\bf A}$. In the sequel we shall
 assume the following time dependence for the moving
 reduced wave vector: \ber \kt_x(t) & = & a \cos(\omega t)+ c, \label{ac}\\ \kt_y(t)
 & = & b \sin(\omega t), \label{bd}\enr
 with constants $a$, $b$ and $c$ and a period of $T=2\pi/\omega$.

 \section{Geometric Phase of the Wave Function Component}\label{GP}
 Solutions of the time dependent \SE,  written for a general form of a periodic $2\times2$ Hamiltonian as
 in \er{HK} and
 valid in the adiabatic limit, were given in Refs.~\onlinecite{EnglmanV1}-\onlinecite{EnglmanV2}, with  special regard to the geometric
 phase acquired near an "orthogonality time" $t_v$. This is the instant at which the initially engaged component of
 the quasi-spinor vanishes, or the state becomes orthogonal to the initial state. There may be several such instants,
  depending on the path taken by the time dependent parameters. As already remarked in the Introduction, the solutions
  of the above cited works went one approximation (in the adiabatic parameter, to be defined shortly) beyond the formal
   adiabatic wave function.
   The instantaneous energy and twice the mixing angle appropriate to the Hamiltonian
   in \er{HK} (and analogously to that in \er{monoham}) are
   \begin{align}
   W(t)&\equiv\sqrt{U({\bf \kt}(t))^2+V({\bf \kt}(t))^2},\\
   \chi(t)&\equiv \arctan\frac{V({\bf \kt}(t))}{U({\bf \kt}(t))}.\label{Wchi}
   \end{align}
    The (small) adiabaticity parameter is in terms of these \beq
\epsilon (t)=\frac{{\dot\chi}(t)}{W(t)}\label{eps}.\enq

We now turn to the solution of the time dependent \SE~ in the form
\beq i\frac{\partial}{\partial t}\left
 ( \begin{array}{cc}f(t)\\g(t)
 \end{array} \right)=
 \left( \begin{array}{cc}
   -U(t) &  V(t) \\
   V(t)  &  U(t)
 \end{array} \right)\left
 ( \begin{array}{cc}f(t)\\g(t)
 \end{array} \right)
 \label{se} \enq
 with the initial condition for an energy eigenstate
 $f(t=0)=1$, $g(t=0)=0$.
In Refs.~\onlinecite{EnglmanV1}-\onlinecite{EnglmanV2} we have obtained (after the
removal of the dynamic phase factor) the following component
amplitudes of the wave function in the neighborhood of an
orthogonality point $t_v$ as
\begin{align} f(t)= & -\frac{1}{2}{\dot
\chi}(t_v)[(t-t_v)\nonumber\\
& - \frac{i}{2|W(t_v)|}(1-\pi
n'(t_v)e^{-2i|W(t_v)|(t-t_v)})]\nonumber\\
 & +  \frac{1}{2}\pi\phi(t_v)\epsilon(t_v),\label{ftv}\\
g(t) = & -\sin \frac{1}{2}\chi(t_v)= - 1.\label{gtv}\end{align} The expressions are correct to the
order of $|{\dot \chi}(t_v)(t-t_v)|^2$. \beq n'(t_v)\equiv\pi^{-1}\frac{\epsilon (0)}{\epsilon
(t_v)}\label{ntag}\enq which ratio compares the slownesses (or adiabaticities, defined in
equation~(\ref{eps})) at the starting points and at the orthogonality point. It is a key quantity in
the phase acquisition phenomenon, as we shall describe below. In contrast, the last term on the right
of \er{ftv} is of little importance. (This has been confirmed both theoretically and by numerical
computation). In it $\phi(t_v)$ is the fractional part of the energy integral $2\int_0^{t_v}W(t)dt$.
(See Eq.~(31) in reference~\onlinecite{VertesiE}.) The energy (or frequency) $2|W(t_v)|$ in the
complex exponential is the energy separation between the two eigenstates near the orthogonality point
and will show up in our treatment of the electronic motion (or current) in section~\ref{ZB} as a
Zitterbewegung.

We shall now describe the geometric phase acquisition process in
terms of the minority component $f(t)$ in equation~(\ref{ftv}). Were it only
the first term in $f(t)$, a phase of $-\pi$ would accrue upon
traversing  the zero at $t=t_v$ from $t<t_v$ to $t>t_v$. This
phase is continuously carried in $f(t)$ and would remain with
the wave function also when a full revolution is made at which
instant the $g(t)$ component becomes again zero.  Thus, $-\pi$ is
indeed the final result (the Berry phase) as long as $n'(t_v)$ is
smaller than or of the order of $1$. However, when $n'(t_v)$ is
numerically large, there are further zeros in the complex t-plane,
which are all located in the lower half of the complex
t-plane (at shallow depths of the order $1/|W(t_v)|$)
\cite{EnglmanV1,VertesiE,EnglmanV2}. Looking now for the moment at
the behavior of the real part of $f(t)$ [$\mathrm{Re}\, f(t)$], with
algebraically increasing  real t, as this passes each adjacent
(complex) zero, $\mathrm{Re}\, f(t)$ acquires a further $-\pi$ phase. This is
also how the full function $f(t)$ acquires its phase, with the role of
the imaginary part being to ensure that the phase of $f(t)$
changes in a smooth, continuous way. The rate  of $\pi$-
acquisitions is clearly $2|W(t_v)|/\pi$ (a constant for a given
orthogonality point $t_v$).

 How many
(complex) zeros are there in the asymptotic limit of large $
|n'(t_v)|$? Taking into account that the modulus of the complex
exponential in equation~(\ref{ftv}) is unity, we clearly see that the complex
zeros arise only as long as $|t-t_v|\leq \frac{\pi
n'(t_v)}{2|W(t_v)|}$, since otherwise the first term will dominate
the function and  zeros are not possible. Taking into account the
previously obtained value $ |2W(t_v)|/\pi$ for the rate of
acquisition of $\pi$'s, we see that the number $n(t_v)$ (an
integer!)  of acquired $\pi$ phases is for either positive or
 negative $t-t_v$'s about $2n'(t_v)$. [The same result would be
 obtained by counting the number of loops around the origin of
 $f(t)$, since each (anti-clockwise) loop contributes a $2\pi$
 phase and each loop around the origin implicates ${\it two}$
 zeros of $\mathrm{Re}\, f(t)$ (as well as of $\mathrm{Im}\, f(t)$).]

The net result is then that the acquired Berry phase around
$t=t_v$ is \beq[1+2n(t_v)]\pi\label{accphase}\enq where $n(t_v)$
is a signed integer (or zero) close to the ratio $n'(t_v)$ defined
in equation~(\ref{eps}). $n'$ is named the inverse slowness ratio and examples
taken from molecular degeneracies have shown that it is unity for
concentric circular motion around a single intersection point
(and, in the case of several intersection points, for concentric
circular motion  around one intersection point provided the motion
is sufficiently far away from other intersections).
 However it  can be large for circling  that {\it starts} close to an intersection point
  and is not concentric with it \cite{EnglmanV1,VertesiE,EnglmanV2,EnglmanYV}.
  For graphene layers similar results for the phase have now
  been found  and are shown in a set of figures. These are obtained by solving the
   time dependent Schr\"{o}dinger equation and in them the acquired phases agree
   very closely  with $(2n'+1)\pi$, in which $n'$ is given by the formula in
   equation~(\ref{ntag}). This depends only on quantities contained in the
   Hamiltonian.
   The case of (monolayer) graphene, with essentially isolated DP's (the ${\bf K}$ and ${\bf K'}$
   points), is formally identical to the single degeneracy results
   shown in section 3 of reference~\onlinecite{EnglmanYV} and will not be reproduced
   here although  monolayer graphene may be the more convenient candidate for the
   experimental verification of the theory. (However, the one order
   of magnitude lower value of $W(t_v)$ in bilayer graphene than
   in monolayer may make the oscillations easier to monitor.)

\begin{figure}
\centerline{\epsfxsize 3.0in \epsffile{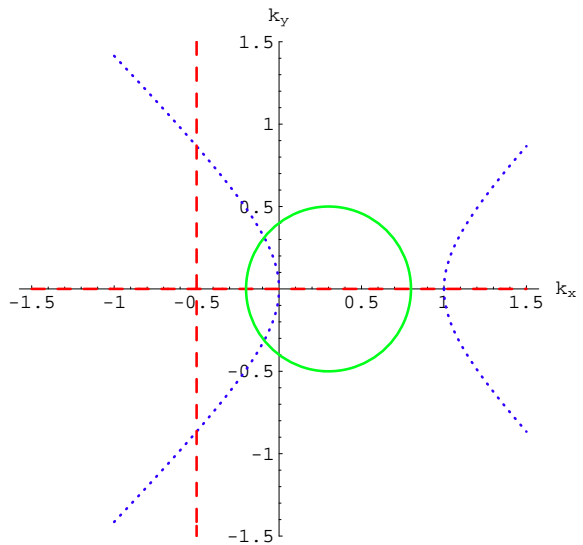}} \caption{(color online). Contours for
bilayer graphene (in the two state-representation). Broken curves (in red) : The potential $V=0$ (in
the Hamiltonian of equation~(\ref{HK})). Dotted curve (in blue) : The potential $U=0$. The circle (in
green) denotes an evolution path. The positions of the  Dirac points  are at the crossings of the
broken and dotted  curves. In the text, the evolution starts at a positive  value of the abscissa.
Orthogonality points $t_v$ , where $\chi=\pi$, are where the path intersects the broken (red) curve.
For the evolution path shown (circling radius =0.5, shift =0.3), there is only a single orthogonal
position $t_v$ on the circular path, which is located on the negative abscissa. For circles shifted to
the left it is possible to have two more $t_v$'s.} \label{fig1}
\end{figure}

     In a graphene bilayer we have a DP at each ${\bf K}$ (and  ${\bf
     K'}$) with three satellite DP's, trigonally situated around
     each $K$-type point. Thus, this is a new situation; also for the reason that there exist
     for it both two and  four dimensional Hilbert space descriptions \cite{KoshinoA,CsertiCD}.
     The Hamiltonian in the former description is shown in \er{U} and \er{V}; some essential
     features of the energy contours are shown in Fig.~\ref{fig1}.

\begin{figure}
\centerline{\epsfxsize 3.5in \epsffile{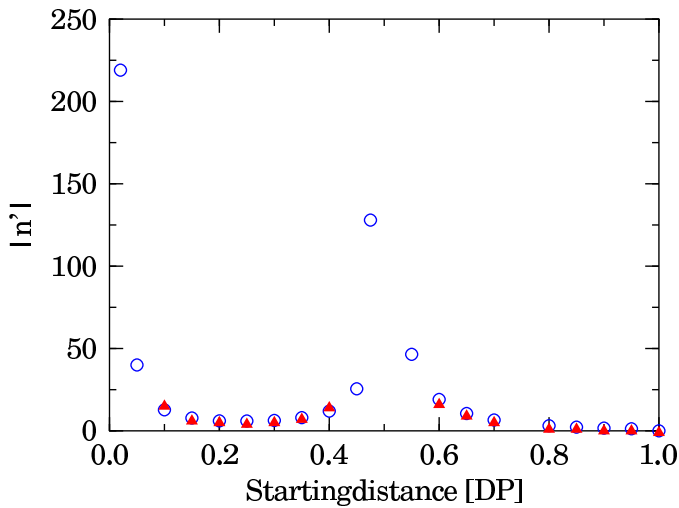}}
\caption{(color online).
Inverse relative slowness parameter $n'$ for  adiabatic circling plotted by open circles (in blue)
against the starting distance from the central Dirac point (DP) in the direction
of a trigonal DP (given in units [DP], measuring the distance of the
trigonal DP from the central one).  $n'$ is approximated  by the
integer $ n$ in the  Berry phase $(2n+1)\pi$ acquired at an
orthogonality point $t_v$ during circling. A circular path with
radius  0.5[DP] was assumed. The inverse slowness parameter $n'$
is large when the path  starts near the central DP or
(numerically) at just
  half way to the trigonal DP. (Here $n'$ turns negative.)
  The triangles (in red) show the results of numerically
  computing the phase jumps. Some triangles are missing because of
  numerical difficulties in achieving convergence.}
\label{fig2}
\end{figure}

     We begin by showing results for  a contour that starts close to and just outside a trigonal
  DP and
  makes a bee-line with a path ultimately skirting round the central DP. [In Koshino and Ando's
  reference~\onlinecite{KoshinoA} figure~2, the
    path might start at an energy of about (or smaller than) $0.5e_0$ (roughly equivalent to an electron density below
    $4$x$10^{10}$cm$^{-2})$  or, otherwise, at $k_x\leq 1.2k_0$, $k_y=0$ and then
    return so as to
    come down beyond the origin of $\bf k$, the $\bf K$-point.] A full contour of this type is not expected
    to change the sign of the wave-function (with the dynamic phase disregarded), since two DP's are
    encompassed by it. In this case one obtains the variation of the $2n$ part of the phase near an orthogonality point,
     which is of interest.
   In figure~\ref{fig2} we show (with open circle symbols) the resulting $n'$ in a rather simple circling situation, when the contour starts to the
   right of the central DP and cycles anti-clockwise round it and inside the trigonal DP's.
   An orthogonality
    point $t_v$ is met at about (but not at exactly) half a full circle. Here the (open path) geometric phase makes a jump,
    whose magnitude is close to the ratio $n'(t_v)$ defined above. For a circular contour around the central DP (which is
    the path here taken) the magnitude of $n'$ depends on the distance of the starting point (located between the rightmost
    and the central DP's). As figure~\ref{fig2} shows, $|n'|$ is large for contours starting close to the central DP and decreases
    as the starting point recedes only to become again large when the starting point is precisely at half way to a
    trigonal DP. This finding arises from the presence of several DP and is absent for a Brillouin zone with
     a single DP. The triangles in the same figure show, for comparison, the
     phase jumps as obtained by numerically solving the time
     dependent \SE~ (the subject of the next subsection).

\section{Numerical Verification of the "Inverse Relative Slowness" Prediction}
 In Refs.~\onlinecite{EnglmanV1}-\onlinecite{EnglmanV2} the Berry-phase was associated  (through what we regard as the proper
 treatment of the time dependent Schr\"{o}dinger equation (TDSE) in the near adiabatic limit) with the quantity $(2n'+1)\pi$ (or $2n'\pi$),
  $n'$ being (as described above)
 the inverse relative slowness. In the quoted articles the theory was verified by numerical solution of the TDSE
 in the adiabatic limit, in the context of nuclear motion in electronically degenerate
 molecules. In the present graphene context (which differs from the molecular setting in essential physical,
 and some formal, details) the numerically obtained phase jump values $n$  have been plotted in Fig.~\ref{fig2} with
 triangles. These
show good agreement with the values (open circles) obtained
 algebraically. Where the triangles are missing, this is  due to
  numerical difficulties in achieving convergence.

\begin{figure}
\centerline{\epsfxsize 3.0in \epsffile{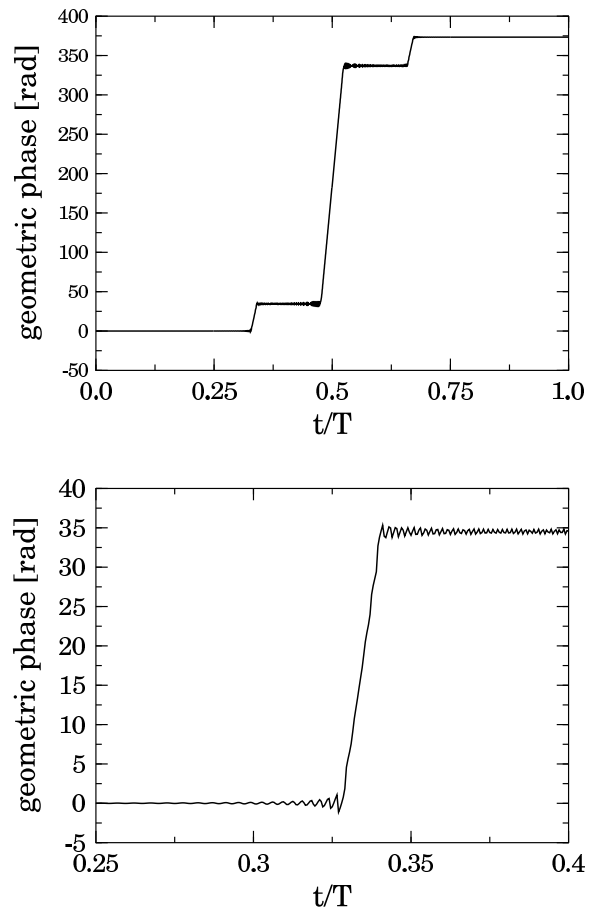}}
\caption{
Adiabatic phase evolution in bilayered graphene (obtained
from numerical solution of the time dependent \SE~within the
two-state formalism). The adiabatic revolution period $T$ is $2\pi
\times 10^{5}$ in inverse units of the Fermi velocity times the DP
distance. Circular cycling meets three  orthogonality points along
its path (as explained in the caption of Fig.~\ref{fig1}). In the upper
drawing, at these points steep rises take place in the phase, with
the middle one (on the left-part of the abscissa in figure~\ref{fig1})
being the longest rise. The near-horizontal wavy part becomes
straightened out, as the motion becomes more adiabatic. In the
lower drawing the first $t_v$ region (above the $k_x$ axis) is
enlarged.}
\label{fig3}
\end{figure}

 An analogous  verification has been performed  by solving the TDSE for a graphene bilayer
 Hamiltonian, including a periodic, time dependent electric field $E(t)$ and calculating at each instant $t$ the phase of the wave function.
In figure~\ref{fig3}, the phase development is shown for one full cycle of
the electric field and the detail for the first step.
The non-constant, rising parts of the phase near the orthogonality
points will be the subject of the section on "Currents".

\begin{figure}
\centerline{\epsfxsize 3.0in \epsffile{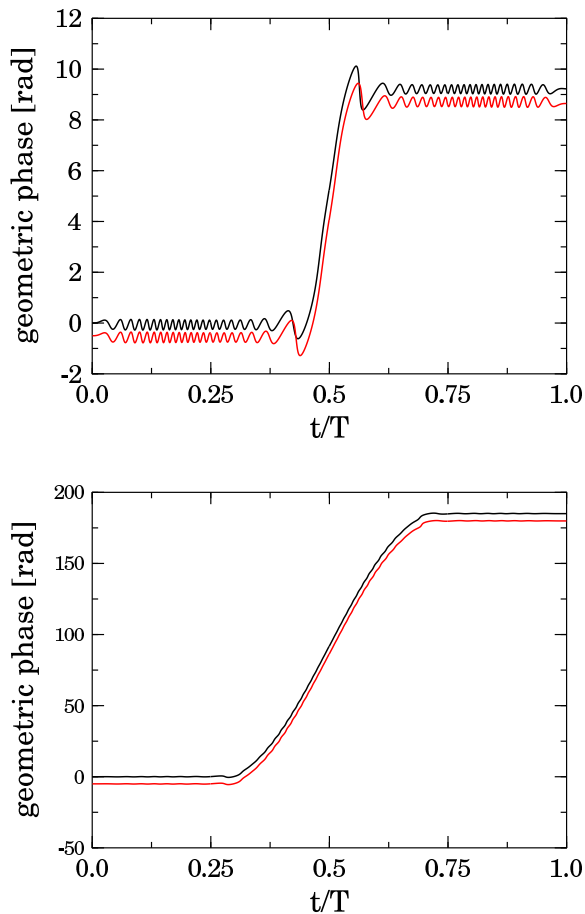}}
\caption{(color online).
Comparison of the Berry phase behaviors of
the four- and two-state models in graphene bilayer. Curves for the
four state model are slightly displaced downward, for clarity. Two
different starting positions to the right of the central Dirac
point (DP) were chosen, expressed in units of the distance to the
right-most DP . Upper frame  starting position: $0.8$. Lower frame
starting position: $0.25$. The adiabatic revolution period $T$ is
$2\pi\times 10^{4}$ in inverse units of the Fermi velocity times the DP
distance. Note the different ordinate scales.}
\label{fig4}
\end{figure}

 The drawings in figures~\ref{fig1}-\ref{fig3} have been based on the two-state (approximate) description of bilayer graphene.
  In figure~\ref{fig4} we compare the component phases obtained from
 forward integration of the time dependent Schr\"{o}dinger equation for the four-state
 with that for the approximated two-state description~\cite
 {KoshinoA}. There are no differences visible to the naked eye,
 which we interpret as  a satisfactory test for the robustness of the
 adiabatic theory formulated in our previous papers.

 \section{Zitterbewegung (ZB) of the Current}\label{ZB}
The equation of motion for the Heisenberg position operator ${\bf
r}(t)$ of an electron is $ i{\dot{\bf r}}(t)=-[H,{\bf r}(t)]$. The
current is ${\bf j}(t)=-e{\dot{\bf r}}(t)$. We shall evaluate this
for a monolayer graphene near its ${\bf K}$ point where the
expressions are quite simple, rather than for a graphene bilayer,
for which the results are analogous, but more complicated.
Inserting the Hamiltonian from \er{monoham} one obtains
immediately the velocity components as \beq {\dot{\bf r}}(t)\equiv
{\dot x}(t){\hat{\bf i}}+ {\dot y}(t){\hat{\bf
j}}=v_F\left(\begin{array}{cc}
   {\hat{\bf i}} &  {\hat{\bf j}} \\
   {\hat{\bf j}}  &  -{\hat{\bf i}}\end{array} \right).\label {vel}\enq
   The observable, expectation value of the velocity is the
   expectation value over the wave function, namely,
   \beq <{\dot{\bf r}}(t)>\equiv v_F (f^*,g^*)\left(\begin{array}{cc}
   {\hat{\bf i}} &  {\hat{\bf j}} \\
   {\hat{\bf j}}  &  -{\hat{\bf i}}\end{array} \right)\left(\begin{array}{cc}
   f  \\
   g\end{array} \right),\label{Ev}\enq where the stars denote
   complex conjugates. Two situations need now to be considered.

   \subsection{Electronic motion during most of the contour path}
\subsubsection{Extreme adiabatic limit, $\epsilon\to 0$}
   Here \beq (f^*,g^*)=(f,g)=(\cos\frac{1}{2}\chi(t),
-\sin\frac{1}{2}\chi(t))\label{fg1}\enq in terms of the mixing angle $\frac{1}{2}\chi(t)$, which
effects the diagonalization of the $2$ x $2$ Hamiltonian. For the monolayer Hamiltonian in
\er{monoham} this is defined through $\chi(t)\equiv \arctan\frac{V}{U}=-\arctan\frac{k_x(t)}{k_y(t)}$,
so that it is
 possible to express the momentum components as $k_x(t)=|{\bf k}|\cos \chi(t)$, $k_y(t)=-|{\bf k}|\sin
 \chi(t)$,  where the time dependence
 arises from the applied vector potential as in equation~(\ref{k(t)}). Then \er{Ev} evaluates to \beq <{\dot{\bf r}}(t)>=
v_F [\cos\chi(t){\hat{\bf i}}- \sin\chi(t){\hat{\bf
j}}],\label{vel2}\enq which merely reaffirms the equivalence of the
time behaviors of the momentum and velocity vectors for a
completely adiabatic motion along most of the pathway.
\subsubsection{First order adiabatic correction}
When the wave function components are expanded correct to the
order of the small adiabatic parameters $\epsilon(t)$, one obtains
the terms describing the ZB of the electronic velocities. \ber
<{\dot x}(t)>_{ZB} & = -\frac{v_F\epsilon(0)}{2}\sin\chi(t)\sin
2I(t),\nonumber\\<{\dot y}(t)>_{ZB} & =
-\frac{v_F\epsilon(0)}{2}\cos\chi(t)\sin 2I(t),\label{velcorr}\enr
where we have denoted the integral over energy as \beq
I(t)\equiv\int ^t_0dt'W(t').\label {It}\enq There may be some
practical difficulties in the  observation of this motion as it is
superimposed on a larger motion shown in equation (\ref{vel2}).

\subsection{Near the orthogonality point $t_v$}\label{near}
Here the eigenstate component amplitudes  $ (f,g)$ take the
time-dependent forms shown in \er{ftv} and equation~(\ref{gtv}). Substitution
into equation~(\ref{Ev}) shows that in the adiabatic limit the $x$-component
moves with a uniform speed of $v_F$. However, the $y$-component of
the electron has the value \beq {\dot y}(t) =\pm 2v_f Re
f(t).\label{yvel}\enq As discussed in section~\ref{GP}, this is a small
quantity, of the order of the small adiabaticity parameter
$\epsilon(t_v)$ which oscillates with the period characteristic of
ZB, given in the present case by $\pi/W(t_v)$. The eigenstate
oscillations take place for a time interval of about
$\frac{n'}{W(t_v)}$, placed symmetrically about $t_v$. This
interval can be quite large in case of strongly non-concentric
circling, for which the relative inverse slowness $n'$ (defined in
equation~(\ref{ntag})) has large values (e.g., of the order of $10^2$)\cite
{comment}.
\section{Subtraction of the Dynamic Phase}\label{dyn}
Results in the previous sections have the following implication
(again phrased for simplicity's sake for the monolayer graphene
case, section~\ref{mono}): For a given radius of circling $k_c$ around a
${\bf K}$-point the acquired  Berry phase can be increased  by
diminishing the distance $\Delta k$ between the starting point and
the center of the circular path. The increase in the Berry phase
is about $2\pi$ times $ n'$ (the relative inverse slowness) which
is unity when the distance $\Delta k=k_c$ and increases beyond
bounds as $\Delta k \to 0$. A hypothetical {\it direct}
experimental determination of the augmented phase would then
likely perform two measurements at or close to these two limits.

Since the above theory relates to the Berry phase, while the total
acquired phase includes also the dynamical phase (designated
$\phi_D$), it is purposeful to estimate the latter. It was shown
in Refs.~\onlinecite{EnglmanV1}-\onlinecite{EnglmanV2} that in the adiabatic limit
[equation~(\ref{adiabatic}) above] this is simply given  by the time integral
of the instantaneous energy. Then for a uniform circular motion in
the $ {\bf k}$-plane  \begin{align} &\phi_D (k_c,\Delta
k)\nonumber \\ & =v_F\int_0^{2\pi/\omega} \sqrt{(k_c\cos \omega t)-\Delta k)^2 +
(k_c\sin\omega t)^2} dt. \label{phiD}\end{align} This elementary integral
has the following values in the two limiting cases discussed
above:
\begin{align}
\textrm{concentric~cycling:}\; \phi_D(k_c,k_c) & =  2\pi
\frac{v_F|k_c|}{\omega},\\
\textrm{touching:}\; \phi_D(k_c,0) & =
2\sqrt{2}\frac{v_F|k_c|}{\omega}. \label{PhiD2}\end{align}
These dynamic phase values can be subtracted from the observed total phase to
obtain the Berry-phase. From our direct computations of the
acquired total phase by the wave function, for situations such as
shown in Fig.~\ref{fig3} we find that for large $|n'|$-values the phase is
dominated by the Berry-phase. (The dynamic phases at the three
phase jumps shown in that figure amount to -20, -60, -20 radians,
respectively.)

\section{Conclusion}

This work has focused on two (interrelated) issues in the context
of mono- and bilayered graphene.

 First, (based on our past works on
adiabatic cycling, which are only briefly recapitulated here) we
have shown that large amplitude (and gauge-invariant) phases of
the electronic wave-functions can be generated in two-dimensional
structures of the graphene-type. A conceptual experimental
procedure  involving a uniform time varying electric field has
been outlined, though without our proposing detailed prescription.
The key geometrical element is the non-concentric adiabatic path
in {\bf k}-space around the ${\bf K}$ points.

The second issue is the association of the {\it developmental stage} of the large amplitude Berry
phase (near the  orthogonality points) with the Zitterbewegung of the electronic motion and its
observational possibility. So far little attention has been given to the possibility of experimentally
detecting the Berry phase in the neighborhood of an orthogonality point; though some theoretically
oriented remarks regarding the abrupt nature of this developmental stage of the Berry phase have been
made \cite{SjoqvistH, BaerYE}. These remarks addressed the minimal $\pm\pi$ phase acquisitions near
each orthogonality point, whereas our proposal for the observation of the phase by Zitterbewegung is
for large Berry phases.

\acknowledgements
Thanks are due to Joshua Zak and Yizhak Yacoby  for educating remarks. T.V. has been
supported by a J\'anos Bolyai Grant of the Hungarian Academy of Sciences.

\end{document}